\documentclass[twocolumn,aps,prd,showpacs,preprintnumbers,amssymb]{revtex4}

\usepackage{dcolumn}
\usepackage{bm}

\begin{document}

\title{Vacuum polarization on the spinning circle}
\author{V. A. De Lorenci and E. S. Moreira Jr.}
\affiliation{ Instituto de Ci\^encias -
Universidade Federal de Itajub\'a \\
Av. BPS 1303 Pinheirinho, 37500-903 Itajub\'a, MG -- Brazil \\
(Email addresses: {\tt lorenci@efei.br, moreira@efei.br})}

\date{December 14, 2001}

\begin{abstract}
Vacuum polarization of a massive scalar field in the background
of a two-dimensional version
of a spinning cosmic string is investigated. It is shown that when
the ``radius of the universe''
is such that spacetime is globally
hyperbolic the vacuum fluctuations are
well behaved, diverging though on the ``chronology horizon''.
Naive use of the formulas when spacetime is nonglobally
hyperbolic leads to unphysical results.
It is also pointed out that the set of normal modes
used previously in the literature to address the problem
gives rise to two-point functions
which do not have a Hadamard form,
and therefore are
not physically acceptable. Such normal modes correspond
to a locally (but not globally) Minkowski time,
which appears to be at first sight a natural choice of time
to implement quantization.
\end{abstract}

\maketitle

\label{int}

The study of quantum fields around
cosmic strings is a pertinent issue
since such defects may play a
role in the cosmological scenario \cite{vil94}.
Most of the literature concerns spinless
cosmic strings (see Ref. \cite{vil94} and references therein),
and only a few works have considered quantum
mechanics and quantum field theory around spinning cosmic strings
\cite{ger89,ger90,mat90,jen93,bez97,lor01}.

The locally flat spacetime around an
infinitely thin
spinning cosmic string \cite{vil94} is
characterized by the
line element
\begin{equation}
ds^{2}=(d\tau+Sd\varphi)^{2}-d\rho^{2}
-\alpha^{2}\rho^{2}d\varphi^{2}-d z^{2}
\label{lee}
\end{equation}
and by the identification
$\left(\tau,\rho ,\varphi, z\right)
\sim  \left(\tau,\rho ,\varphi +2\pi,z\right)$,
where $0<\alpha\leq 1$ is the cone parameter
and $S\geq 0$ is the spin density
(clearly the Minkowski spacetime corresponds to
$S=0$ and $\alpha=1$).
As the region for which $\rho<S/\alpha$
contains closed timelike contours
the spacetime is not globally hyperbolic.
In other words a global time is not available,
so that it is not clear whether
quantum theory makes sense in this background \cite{ful89}.

The study of a relativistic quantum scalar particle moving on the
spinning cone [the corresponding three-dimensional
line element is obtained
from Eq. (\ref{lee}) by setting $dz=0$] has shown
that a nonvanishing $S$ spoils unitarity \cite{ger89}.
It has been speculated that this sort of first quantized
pathology could be eliminated
in the second quantized approach. However,
Ref. \cite{lor01} seems to frustrate this possibility
by showing that the vacuum fluctuations of a massless scalar field
diverge on concentric cylindrical shells around
the spinning cosmic string.
These pathological results have been attributed
to the nonglobally hyperbolic nature of the background.

In order to exhibit clearly aspects of global hyperbolicity
(and related issues) in actual calculations, this
work will consider a toy model
which consists of a quantum scalar field existing in
a two dimensional spacetime whose line element is
obtained by truncating Eq. (\ref{lee})  as
\cite{jen93}

\begin{equation}
ds^{2}=(d\tau+S d\varphi)^{2}-\rho^{2}d\varphi^{2}
\label{le}
\end{equation}
($\alpha$ was dropped
since it can be removed by redefining  the parameter $\rho$)
and observing
\begin{equation}
\left(\tau,\varphi\right)
\sim  \left(\tau,\varphi +2\pi\right).
\label{id}
\end{equation}
It is clear that $S=0$ corresponds to a cylindrical
spacetime of periodicity length $2\pi\rho$.
The main pedagogical feature of this toy model is
that, for a given ``radius of the universe''
$\rho>0$,  one can tune  the spin $S$ such that
the background is globally hyperbolic ($\rho>S$)
or otherwise ($\rho\leq S$).

Assuming $\rho>S$ and defining new parameters
\begin{eqnarray}
r:=\sqrt{\rho^{2}-S^{2}},&\hspace{1cm}&
\Omega:=\frac{S}{\rho\sqrt{\rho^{2}-S^{2}}}
\label{nparameters}
\end{eqnarray}
and a new time coordinate
\begin{equation}
t:=\frac{\tau}{\sqrt{1-\Omega^{2}r^{2}}},
\label{ncoordinates}
\end{equation}
it follows that Eq. (\ref{le}) can be recast as
\begin{equation}
ds^{2}=(1-\Omega^{2}r^{2})dt^{2}
+2\Omega r^{2}dt\ d\varphi-r^{2}d\varphi^{2}.
\label{rle}
\end{equation}
Defining further
$\theta:=\varphi-\Omega t$
Eq. (\ref{rle}) becomes
\begin{equation}
ds^{2}=dt^{2}-dx^{2},
\label{mle}
\end{equation}
with $x:=r\theta$, and Eq. (\ref{id}) leads to
\begin{equation}
\left(t,x\right)
\sim  (t,\ x+L),
\label{mid}
\end{equation}
where
\begin{equation}
L:= 2\pi\sqrt{\rho^{2}-S^{2}}.
\label{pl}
\end{equation}
Therefore the spinning circle defined in Eq. (\ref{le})
is a cylindrical spacetime
with the spin $S$ encapsulated in the periodicity length $L$.

Alternatively, Eq. (\ref{rle}) describes the background
as a cylindrical spacetime seen from a rotating frame
with angular velocity $-\Omega$. Noting that $\rho>S$
leads to $r>0$ and consequently $\Omega r<1$,
by requiring global hyperbolicity one ensures that the
velocity of the rotating frame is less than the velocity of
light. It should also be remarked that as $S\rightarrow\rho$,
$r\rightarrow 0$ and $\Omega r\rightarrow 1$. That is, on the
``chronology horizon'' ($\rho=S$) the corresponding
cylindrical spacetime collapses to its axis and  the
associated rotating frame reaches the velocity of light.

As long as $\rho>S$, i.e. the spacetime is globally
hyperbolic, one can implement quantization in any of the
frames considered above
[corresponding to Eqs. (\ref{le}), (\ref{rle}), and (\ref{mle})],
since the time coordinates are
genuine global times (in the sense that they parametrize
Cauchy surfaces) and the corresponding time translation
Killing vectors are globally timelike.
The usual procedure to quantize a scalar field $\phi({\rm x})$
reveals that these frames have identical sets of normal modes,
and therefore identical vacuum states (which
is not surprising since there is
no event horizon involved \cite{let81}).
It is clear from Eqs. (\ref{mle}) and (\ref{mid}) that
the set of normal modes is that
associated with a
cylindrical two-dimensional
spacetime, which is well known in the literature
\cite{dav82,gri94}.

The Hadamard function
corresponding to the field modes
appearing in Eq. (4.1) of Ref. \cite{dav82}
[with $\omega=(k^{2}+m^{2})^{1/2}$]
can be cast as
\begin{widetext}
\begin{equation}
G^{(1)}({\rm x},{\rm x}')=
\frac{1}{mL}
-\frac{1}{4\pi}
\ln\left\{4\left[\cos(2\pi\Delta t/L)-
\cos(2\pi\Delta x/L)\right]^{2}\right\},
\label{hf}
\end{equation}
\end{widetext}
where $\Delta t:=t-t'$, $\Delta x:=x-x'$, and a small mass $m$
was taken into account to prevent the usual infrared divergence
[in fact $mL<<1$ has been considered and, accordingly,
higher powers of $mL$ were omitted in Eq. (\ref{hf})].
Incidentally one may check that the
massless contribution in Eq. (\ref{hf}) reproduces Eq. (4.23)
in Ref. \cite{dav82}.

A quick examination of Eqs. (\ref{pl}) and (\ref{hf}) shows that
$G^{(1)}({\rm x},{\rm x}')$ diverges when $\rho=S$.
As the vacuum fluctuations
$\langle\phi ^{2}({\rm x})\rangle$
can be  formally obtained from
the Hadamard function \cite{dav82,wal94}, i.e.,
\begin{equation}
\langle\phi ^{2}({\rm x})\rangle=
\lim_{{\rm x}'\rightarrow {\rm x}}\frac{1}{2}
G^{(1)}({\rm x},{\rm x}'),
\label{vf}
\end{equation}
one may wonder whether $\langle\phi ^{2}({\rm x})\rangle$ itself
diverges on the ``chronology horizon''. That is indeed the
case as the following calculations show.

As the background is flat, the ultraviolet divergence
arising in Eq. (\ref{vf}) can be cured simply
by removing the contribution in
Minkowski spacetime.
It follows from Eq. (\ref{hf}) that
its short distance behavior is given by
\begin{equation}
G^{(1)}({\rm x},{\rm x}')=
\frac{1}{mL}
-\frac{1}{2\pi}
\ln(4\pi^{2}|\sigma|/L^{2}),
\label{sdhf}
\end{equation}
where $\sigma:=(\Delta t)^{2}-(\Delta x)^{2}$.
The Minkowski contribution is given by \cite{dav82}
\begin{equation}
G^{(1)}_{0}({\rm x},{\rm x}')=
-\frac{1}{2\pi}\left[\ln(m^{2}|\sigma|/4)
+2\gamma\right],
\label{mhf}
\end{equation}
with $\gamma$ denoting the Euler constant.
By subtracting Eq. (\ref{mhf}) from Eq. (\ref{sdhf})
before taking the limit in Eq. (\ref{vf}),
one finds that
\begin{equation}
\langle\phi ^{2}({\rm x})\rangle=
\frac{1}{2mL}
+\frac{1}{2\pi}\left[
\ln(mL/4\pi)+\gamma\right],
\label{rvf}
\end{equation}
where the usual infrared divergence
($m=0$) and the
chronology divergence ($L=0$)
have identical structures.
It should also be pointed out that the vacuum
fluctuations lose reality
when Eq. (\ref{rvf}) is naively used
when $\rho<S$ [cf.  Eq. (\ref{pl})].

The procedure outlined above can be extended to
evaluate vacuum expectation values of other
quantities such as the components of the energy
momentum tensor.
The energy density, the
pressure, and the momentum density are
given, respectively, by
$-\pi/6L^{2}+m/2L$, $-\pi/6L^{2}$, and $0$,
with respect to the frame corresponding to Eq. (\ref{mle})
\cite{dav82,gri94}.
By evoking Eq. (\ref{pl}) one sees that the energy momentum
tensor also diverges on the ``chronology horizon''.
(It is  worth remarking that the spin $S$ does not induce
momentum density, which is reasonable since $S\neq 0$
affects $L$ only.)

Considerations on a certain ``time-helical'' structure \cite{des84}
are in order. By redefining the time coordinate according to
\begin{equation}
T:=\tau+S\varphi
\label{ht}
\end{equation}
Eq. (\ref{le}) can be recast in Minkowski
form,
\begin{equation}
ds^{2}=dT^{2}-dX^{2},
\label{htle}
\end{equation}
where $X:=\rho\varphi$.
Observing Eqs. (\ref{id}) and (\ref{ht}),
it follows that
\begin{equation}
\left(T,X\right)
\sim  \left(T+2\pi S,\ X+{\rm L}\right),
\label{htid}
\end{equation}
with ${\rm L}:=2\pi\rho$.
Equations (\ref{htle}) and (\ref{htid})
should be compared with Eqs.
(\ref{mle}) and (\ref{mid}), respectively.
The main difference is that $T$ satisfies
an unusual identification, giving rise to a
``time-helical'' structure.

The time coordinates $T$ and $t$
are related by [cf. Eq. (\ref{ncoordinates})]
\begin{equation}
T=\frac{t-Vx}{\sqrt{1-V^{2}}},
\label{htmt}
\end{equation}
where $V:=-\Omega r$.
If $x$ were to label points on the line Eq. (\ref{htmt})
would be identified as a genuine Lorentz transformation.
Nevertheless,
$x$ labels points on the circle [cf. Eq. (\ref{mid})]
and Eq. (\ref{htmt}) can only be considered as a
Lorentz transformation locally, i.e., when $T$, $t$, and $x$
are replaced by $dT$, $dt$, and $dx$, respectively.
It turns out that (when $V\neq 0$)
the ``helical time''  $T$
is locally Minkowski only, whereas $t$ is globally Minkowski.

The time coordinate $T$ has been used in Ref. \cite{jen93}
to quantize a scalar field on the spinning circle.
Although the background is flat,
the corresponding two-point function
presents (in addition to the usual flat divergence)
short distance divergences
containing  the spin $S$ as a factor.
Such divergences certainly cannot be renormalized
away by subtracting the Minkowski contribution,
and one says that the two-point function does not
have the Hadamard form \cite{wal94}.
This unphysical feature is not surprising
if one recalls that the standard knowledge
requires a global time in implementing quantization.
In fact, $T$ is not a global time,
since constant values of $T$ do not parametrize
Cauchy surfaces as long as $S\neq 0$ [cf. Eq. (\ref{htid})].
Therefore the results in Ref. \cite{jen93}
are spoiled by improper use of $T$ as a global time.
[It should be mentioned that improper use
of ``helical times'' as global times may also spoil results
in other contexts. For instance, the use of $T$ as
given by Eq. (\ref{htmt}) to study the propagation
of light in a  rotating frame
yields results that contradict well established
experimental facts \cite{gro77}.]

Summarizing, this toy model illustrated
in actual calculations the
relevance of global hyperbolicity
for a consistent quantization, and some
consequences of the improper  use of ``helical times''
to address global issues.

\begin{acknowledgments}
The authors are grateful to George Matsas,
Raphael De Paola, Renato Klippert, and Ricardo Medina
for useful discussions.
This work was partially supported by the
Brazilian research council CNPq.
\end{acknowledgments}


\end{document}